\documentclass[final,1p,times]{elsarticleTC}

\usepackage{amsmath,amssymb,amsfonts,amscd}
\usepackage{graphicx}

\usepackage{color,soul}

\begin{document}

\begin{frontmatter}

\title{Quantum graph vertices with minimal number of passbands}
\author
{Sergey S. Poghosyan}
\ead{sergey.poghosyan@kochi-tech.ac.jp}

\author
{Taksu Cheon}
\ead{taksu.cheon@kochi-tech.ac.jp}
\corref{cor1}
\address
{Laboratory of Physics, Kochi University of Technology,
Tosa Yamada, Kochi 782-8502, Japan}
%

\date{\today}

\begin{abstract}
We study a set of scattering matrices of quantum graphs containing minimal number of passbands, i.e., maximal number of zero elements.  The cases of even and odd vertex degree are considered. Using a solution of inverse scattering problem, we reconstruct boundary conditions of scale-invariant vertex couplings. Potential-controlled universal flat filtering properties are found for considered types of vertex couplings.  Obtained boundary conditions are approximated by simple graphs carrying only $\delta$ potentials and inner magnetic field.
\end{abstract}


\end{frontmatter}

\section{Introduction}

%
The interest in the study of quantum single-electron devices has been on the rise for more than a decade. 
The purpose of this paper is to continue an investigation of mathematical aspects
of quantum devices controlling a transmission of electrons through networks. We will consider quantum devices based on quantum graphs with spectral filtering properties.

The quantum graph theory describes a dynamics of quantum particles propagating through graph, which consists of nodes and one-dimensional
lines. Historically, quantum graphs emerged as an approximation of organic molecules in quantum chemistry \cite{orgmol}. Later it started to attract an
attention of scientists as a mathematical model of single-electron devices based on interconnected nanoscale wires composed of carbon,
semiconductors or other materials \cite{nano1,nano2,nano3}.
Various theoretical approaches were applied to investigate scattering properties of quantum graphs \cite{CM13,EM13}.

In the theory of quantum mechanics, it is well known that the one-dimensional line with $\delta$-potential can be used as a high-energy filter, and
$\delta'$-potential works as a low-energy one. More complex spectral branching filters can be designed with singular vertex of 
scale-invariant type (also referred to as F\"ul\"op-Tsutsui type) boundary conditions \cite{FT00}. The significant contribution in revealing new filtering properties of quantum graphs 
has been an introduction of
external controlling potentials. The application of controlling potentials on the lines of quantum graphs gives an opportunity to design a
tunable quantum band-pass spectral filters. Filtering characteristics of the quantum graph vertex strongly depend on strength of potentials
and boundary conditions. The problem of finding scattering matrix for arbitrary number of constant controlling potentials and given %
scale-invariant couplings was solved analytically \cite{Pot-contr}. However, it is still 
an open question how to apply that analytical result in the case of large vertex degree.

Another important issue is a realization of graph vertices with 
scale-invariant type boundary conditions in physical systems. It was shown
that such couplings could be approximated by auxiliary graphs, carrying only $\delta$ potentials and inner magnetic field \cite{Pot-contr}. The scattering matrix of
approximating scheme converges to initial one, when the lengths of edges tend to zero in a certain limit.

The characteristics of interaction between incoming particle and quantum graph vertex are determined from boundary
conditions. Having 
scale-invariant type couplings and no external electromagnetic field, one can obtain energy-independent probabilities
of reflection and transmission through star graph lines. On the other hand, measuring scattering matrix elements from the experiment, we
can reconstruct boundary conditions by solving inverse problem. The necessary and sufficient conditions on arbitrary matrix to be a scattering
matrix of star graph with 
scale-invariant couplings are Hermiticity and unitarity. Investigation of the whole set of Hermitian unitary matrices is
a complicated task. However, it is possible and worthwhile to study special subsets of Hermitian unitary matrices and particularly the class
of modularly permutation-symmetric scattering matrices was examined in \cite{ModPerm}.
The second important case is the set of reflectionless equi-transmitting scattering matrices, which is studied in connection to matching conditions in quantum graphs \cite{Inverse-prob, KOR13}.

In this paper, we study another 
family of Hermitian unitary matrices of finite dimension, i.e., the set of scattering matrices with maximal
number of zeros corresponding to minimal number of passbands. 
It is shown that, for the matrix of even dimension $n$,  such family is described by $\frac{n}{2}-1$ complex parameters,
while for odd dimension $n$, it is completely specified by $\frac{n-1}{2}$ complex numbers.
After reconstruction of scale-invariant boundary conditions by solving
inverse scattering problem, we investigate an influence of external electromagnetic field.
Potential-controlled universal
flat filtering properties are revealed for considered types of vertex couplings. To apply quantum graphs in
physical systems, obtained boundary conditions are approximated by simple graphs carrying only $\delta$ potentials and inner magnetic field.

\section{Scale invariant couplings and scattering matrix}
We consider a scattering problem of  a single particle on a quantum graph node containing $n$ edges and
boundary conditions of F\"ul\"op-Tsutsui type couplings \cite{FT00, FTinter}, which is a  scale invariant subfamily of general boundary condition \cite{KS99}. If we introduce boundary-value vectors
$\Psi$ and $\Psi^\prime$ by
\begin{eqnarray}
\label{e12}
\Psi =  \begin{pmatrix} \psi_1(0) \\ \vdots \\ \psi_n(0) \end{pmatrix} ,
\quad
\Psi^\prime =  \begin{pmatrix} \psi^\prime_1(0) \\ \vdots \\ \psi^\prime_n(0) \end{pmatrix} ,
\end{eqnarray}
then the boundary conditions can be represented by
\begin{eqnarray}
\label{e1}
\begin{pmatrix}I^{(m)} & T \\ 0 & 0 \end{pmatrix} \Psi^\prime
= \begin{pmatrix}0 & 0 \\ -T^\dagger & I^{(n-m)} \end{pmatrix} \Psi ,
\end{eqnarray}
where $m$ is an integer $m=1,2,...,n-1$; $I^{(l)}$ is an identity matrix of dimension $l$ and $T$ is a complex matrix
of size $m \times (n-m)$. Let us consider an incoming quantum particle from the $j$-th edge, which scattered on a vertex.
The scattering wave function on the $i$-th line $\psi_i^{(j)}(x)$ can be presented in the form

\begin{eqnarray}
\label{e14}
\psi_i^{(j)}(x) =  \delta_{i,j} e^{-{\rm i} k x} + {\cal S}_{i,j} e^{{\rm i} k x},
\end{eqnarray}
where ${\cal S}$ is a scattering matrix. We also assumed that the origin of the coordinates correspond to the vertex.
Calculations show that the scattering matrix can be expressed by formula
\begin{eqnarray}
\label{e404}
{\cal S} = -I^{(n)} + 2
 \begin{pmatrix}I^{(m)} \\ T^\dagger  \end{pmatrix}
\left( I^{(m)} + T T^\dagger \right)^{-1}
\begin{pmatrix}I^{(m)}  \,\ T  \end{pmatrix} .
\ \
\end{eqnarray}
On the other hand one can address an inverse scattering problem, i.e., having an explicit form of scattering
matrix ${\cal S}$ to reconstruct boundary conditions (\ref{e1}). Note that arbitrary matrix of dimension $n$ cannot serve
as a scattering matrix of quantum graph. As it was mentioned above, the necessary and sufficient condition on such matrix is that ${\cal S}$
has to be unitary and Hermitian. Having these conditions, we set $m=rank({\cal S}+I^{(n)})$ and divide ${\cal S}$ into blocks
${\cal S}_{11}$, ${\cal S}_{12}$, ${\cal S}_{21}$ and ${\cal S}_{22}$ with dimensions $ m \times m$, $m \times (n-m)$,
$(n-m) \times m$ and $(n-m) \times (n-m)$ respectively,
\begin{eqnarray}
\label{e21}
{\cal S} =\begin{pmatrix}{\cal S}_{11} & {\cal S}_{12} \\ {\cal S}_{21} & {\cal S}_{22} \end{pmatrix} ,
\end{eqnarray}
with 
\begin{eqnarray}
\label{s_to_t}
{\cal S}_{11} &=& -I^{(m)} + 2 \left( I^{(m)} + T T^\dagger \right)^{-1} \nonumber, \\
{\cal S}_{12} &=& {\cal S}_{21}^\dagger = 2 \left( I^{(m)} + T T^\dagger \right)^{-1} T, \nonumber \\
{\cal S}_{22} &=& I^{(n-m)} - 2 \left( I^{(n-m)} + T^\dagger T \right)^{-1}. 
\end{eqnarray}
Changing a numbering of edges to make $I^{(m)}+{\cal S}_{11}$ regular if it is necessary, from equations (\ref{s_to_t})
we obtain a matrix $T$
\begin{eqnarray}
\label{e25}
T = \left(  I^{(m)} + {\cal S}_{11}  \right)^{-1} {\cal S}_{12}
 = {\cal S}_{21}^\dagger \left(  I^{(n-m) } - {\cal S}_{22} \right)^{-1}  .
 \quad
\end{eqnarray}
We will use formula (\ref{e25}) for scattering matrices with maximal number of zero elements to reconstruct
boundary conditions (\ref{e1}). Having an explicit form of the matrix $T$, one can investigate filtering properties of the star
graph with controlling potentials on edges.

Suppose the constant potentials $V_1,\ldots,V_n$ are applied on graph edges. Then the new scattering matrix,
which depends also on particle's energy $E$, is given by
\begin{equation}\label{SFTpot}
{\cal S}(E;V_1,\ldots,V_n)=
-I^{(n)}+2\left(\begin{array}{c}
Q_{(1)} \\
Q_{(2)}T^\dagger
\end{array}\right)
\left(Q_{(1)}^2+TQ_{(2)}^2T^\dagger\right)^{-1}
\left(\begin{array}{cc}
Q_{(1)} & TQ_{(2)}
\end{array}\right)\,,
\end{equation}
where
\begin{equation}\label{SFTQpot}
Q_{(1)}=diag\left(\sqrt[4]{1-\frac{V_1}{E}},\ldots,\sqrt[4]{1-\frac{V_m}{E}}\right)\,,\quad Q_{(2)}=diag\left(\sqrt[4]{1-\frac{V_{m+1}}{E}},\ldots,\sqrt[4]{1-\frac{V_n}{E}}\right)\,.
\end{equation}
The formula (\ref{SFTpot}) generalizes (\ref{e404}) and can be obtained directly by matrix
computations \cite{Pot-contr}.

\section{Block scattering matrices of quantum star graphs with even vertex degree}

In this paper we address the problem of finding classes of boundary conditions of quantum graphs corresponding to scattering matrices with
maximal number of zero elements. We will consider only completely connected edges, which means that
the star graph cannot be divided into two or more parts in such way that the whole scattering problem effectively could be considered
separately on that subgraphs. We will see that such family of matrices can be divided into two sets, depending on
parity of dimension. In this section we consider a family of matrices of $n=2r$ dimension of the form

\begin{equation}\label{S_even}
{\cal S}^{(2r)}=
\left(\begin{array}{cc}
0 &  A \\
A^\dagger & 0
\end{array}\right).
\end{equation}
where $A$ is a unitary matrix of size $r\times r$. 
First of all, notice that for arbitrary unitary matrix $A$, the matrix (\ref{S_even}) is both Hermitian and unitary.
Hence, an arbitrary matrix of the form (\ref{S_even})
can serve as a scattering matrix of the scattering problem with corresponding boundary conditions of scale invariant
couplings (\ref{e1}). 
Such block scattering matrices (\ref{S_even}) have been first considered in paper \cite{Pot-contr}
for the case $r=2$.

To reconstruct boundary conditions, let us find a spectrum of operator (\ref{S_even}).
It is obvious that the set of eigenvalues of ${\cal S}^{(2r)}$ consists of either $1$ or $-1$ since the matrix
${\cal S}^{(2r)}$ is Hermitian and unitary. Let us consider one of the eigenvectors of the ${\cal S}$-matrix, having the form
$\left( \begin{array}{cc} \Psi_1 \\ \Psi_2 \end{array} \right)$, where $\Psi_1$ and $\Psi_2$
are columns with lengths $r$, and denote the corresponding eigenvalue by $\lambda\; (\lambda=\pm 1)$.
Notice that the vector $\left( \begin{array}{cc} -\Psi_1 \\ \Psi_2 \end{array} \right)$
is also an eigenvector of ${\cal S}^{(2r)}$ and the corresponding eigenvalue has an opposite sign to $\lambda$.
Therefore the half part of eigenvalues are $1$, the other part are $-1$, and in boundary conditions (\ref{e1}) we have $m=r$.
Applying the formula (\ref{e25}) for ${\cal S}_{11}=0, {\cal S}_{12}=A$, we obtain $T=A$. Thus, our boundary conditions
take a form
\begin{eqnarray}
\label{e1N}
\begin{pmatrix}I^{(r)} & A \\ 0 & 0 \end{pmatrix} \Psi^\prime
= \begin{pmatrix}0 & 0 \\ -A^\dagger & I^{(r)} \end{pmatrix} \Psi .
\end{eqnarray}

Now, let us add a constant potential $U$ on one of the edges of the star graph. To obtain a new ${\cal S}$-matrix, which depends also on a total energy $E$,
we use formulas (\ref{SFTpot}) and (\ref{SFTQpot}) with assumptions $V_i=0, (i=1,...,2r-1)$ and $V_{2r}=U$. After some computations it is not hard to
show that for the energies larger than potential height $E>U$ the scattering matrix takes a form

\begin{eqnarray}
\label{SNE}
{\cal S}=\begin{pmatrix}A_{1r}A_{1r}^*f_1(E) & A_{1r}A_{2r}^*f_1(E) &  ... & A_{1r}A_{rr}^*f_1(E) & A_{11} & ... & A_{1,r-1} & A_{1r} f_2(E) \\
A_{2r}A_{1r}^*f_1(E) & A_{2r}A_{2r}^*f_1(E) &  ... & A_{2r}A_{rr}^*f_1(E) & A_{21} & ... & A_{2,r-1} & A_{2r} f_2(E) \\
\vdots & \vdots &  & \vdots & \vdots &  & \vdots & \vdots \\
A_{rr}A_{1r}^*f_1(E) & A_{rr}A_{2r}^*f_1(E) &  ... & A_{rr}A_{rr}^*f_1(E) & A_{r1} & ... & A_{r,r-1} & A_{rr} f_2(E) \\
A_{11}^* & A_{21}^* & ... & A_{r1}^* & 0 & ... & 0 & 0 \\
A_{12}^* & A_{22}^* & ... & A_{r2}^* & 0 & ... & 0 & 0 \\
\vdots & \vdots &  & \vdots & \vdots &  & \vdots & \vdots \\
A_{1, r-1}^* & A_{2,r-1}^* & ... & A_{r, r-1}^* & 0 & ... & 0 & 0 \\
A_{1r}^* f_2(E) & A_{2r}^* f_2(E) & ... & A_{rr}^* f_2(E) & 0 & ... & 0 & -f_1(E)
\end{pmatrix},
\end{eqnarray}
where the energy dependent functions are defined by
\begin{eqnarray}
f_1(E)=\frac{1-\sqrt{1-\frac{U}{E}}}{1+\sqrt{1-\frac{U}{E}}}, \\
f_2(E)=\frac{2\sqrt{1-\frac{U}{E}}}{1+\sqrt{1-\frac{U}{E}}}.
\end{eqnarray}

\begin{figure}[ht]\center
\begin{tabular}{cc}
\includegraphics[width=43mm]{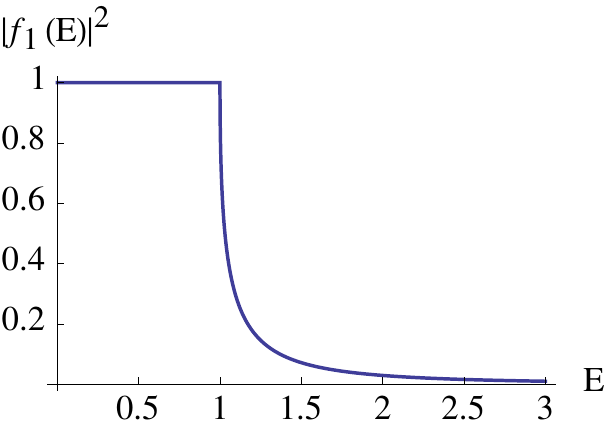}
&
\includegraphics[width=43mm]{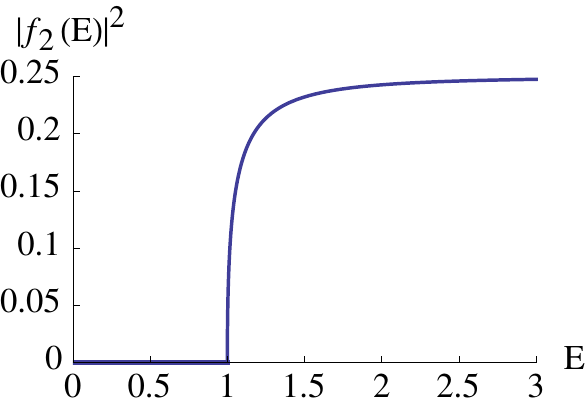} \\
(a)
&
(b)
\end{tabular}
\caption{Energy dependence of functions $|f_1(E)|^2$ (a) and $|f_2(E)|^2$ (b) in the case $U=1$. }
\label{fig1}
\end{figure}

For the case $E<U$ the ${\cal S}$-matrix elements are the same as in (\ref{SNE}), except the last row and last column.
When the energy is less than applied potential, we have an exponential decay of wave function,
so the last row and last column of the matrix (\ref{SNE}) turn into zero. We will assume  $f_2(E)=0$ for $E<U$.
The probability that the particle passes into the $i$-th edge if it initially came from the $j$-th one, is
$\left| S_{ij} \right|^2$. In figure \ref{fig1} the energy dependence of functions
$\left| f_{1}(E) \right|^2$ and $\left| f_{2}(E) \right|^2$ is shown in the case $U=1$. It can be noticed that the absolute values of elements of scattering
matrix in the low energy case $E<U$ are constants. Such universal dependence of passing probabilities was initially obtained in \cite{Pot-contr}.

\section{Block scattering matrices with maximal number of zeros} \label{sec3}

In this section we will consider particular subset of ${\cal S}$-matrices of even dimension with maximal number of zero elements.
The matrix (\ref{S_even}) contains large number of zero elements. However, the number of zeros can be increased by choosing particular form
of unitary matrix $A$. Also from formula (\ref{SNE}) it can be noticed that if we choose $A$ matrix with several zero elements, the corresponding passing amplitudes
will stay zero even after applying a constant potential. Using a condition of completely connected star graph, it is easy to show that in the case $r=3$ the $A$ block of
corresponding scattering matrix can contain no more than one zero. An example of such matrix is

\begin{eqnarray}
\label{A3}
A^{(3)}=
\begin{pmatrix}
0 & \frac{\kappa_1}{\sqrt{1+|\kappa_1|^2}} & \frac{-1}{\sqrt{1+|\kappa_1|^2}} \\
 \frac{\kappa_2^*}{\sqrt{1+| \kappa_2|^2}} &  \frac{1}{\sqrt{1+|\kappa_1|^2}\sqrt{1+|\kappa_2|^2}} &  \frac{\kappa_1^*}{\sqrt{1+|\kappa_1|^2}\sqrt{1+|\kappa_2|^2}} \\
 \frac{-1}{\sqrt{1+| \kappa_2|^2}} &   \frac{\kappa_2}{\sqrt{1+|\kappa_1|^2}\sqrt{1+|\kappa_2|^2}} &  \frac{\kappa_1^* \kappa_2}{\sqrt{1+|\kappa_1|^2}\sqrt{1+|\kappa_2|^2}}
\end{pmatrix},
\end{eqnarray}
where $\kappa_1$ and $\kappa_2$ are arbitrary non-zero complex constants. Other unitary matrices containing one zero element,
can be obtained from (\ref{A3}) by interchanging rows or columns and taking particular values of $\kappa_1$ and $\kappa_2$.
To obtain $A$ matrices of higher dimensions, we generalize a construction (\ref{A3}) by introducing new notations

\begin{equation}
\label{ker}
\begin{array}{c}
Q_1({\kappa})=\begin{pmatrix}
\frac{\kappa}{\sqrt{1+|\kappa|^2}} & \frac{1}{\sqrt{1+|\kappa|^2}}
\end{pmatrix},
\\
\\
Q_2({\kappa})=\begin{pmatrix}
\frac{1}{\sqrt{1+|\kappa|^2}} & \frac{-\kappa^*}{\sqrt{1+|\kappa|^2}}
\end{pmatrix},
\\
\\
K_1({\kappa_1,\kappa_2})=
\begin{pmatrix}
\frac{-1}{\sqrt{1+|\kappa_1|^2}\sqrt{1+|\kappa_2|^2}} &  \frac{\kappa_1^*}{\sqrt{1+|\kappa_1|^2}\sqrt{1+|\kappa_2|^2}} \\
\frac{\kappa_2}{\sqrt{1+|\kappa_1|^2}\sqrt{1+|\kappa_2|^2}} &  \frac{-\kappa_1^* \kappa_2}{\sqrt{1+|\kappa_1|^2}\sqrt{1+|\kappa_2|^2}}
\end{pmatrix},
\\
\\
K_2({\kappa_1,\kappa_2})=
\begin{pmatrix}
\frac{\kappa_1^*\kappa_2}{\sqrt{1+|\kappa_1|^2}\sqrt{1+|\kappa_2|^2}} &  \frac{\kappa_1^*}{\sqrt{1+|\kappa_1|^2}\sqrt{1+|\kappa_2|^2}} \\
\frac{\kappa_2}{\sqrt{1+|\kappa_1|^2}\sqrt{1+|\kappa_2|^2}} &  \frac{1}{\sqrt{1+|\kappa_1|^2}\sqrt{1+|\kappa_2|^2}}
\end{pmatrix},

\end{array}
\end{equation}

For odd $r=2p+1$ our generalized matrix can be represented by

\begin{eqnarray}
\label{ANodd}
A^{(2p+1)}=
\begin{pmatrix}
0 & 0 & \ldots & 0 & 0 & Q_1(\kappa_1) \\
0 & 0 & \ldots & 0 & K_2(\kappa_2, \kappa_3) & K_1(\kappa_1, \kappa_2) \\
0 & 0 & \ldots & K_2(\kappa_4, \kappa_5) & K_1(\kappa_3, \kappa_4) & 0 \\
\vdots & \vdots &  & \vdots & \vdots & \vdots  \\
0 & K_2(\kappa_{r-3}, \kappa_{r-2}) & \ldots & 0 & 0 & 0 \\
Q_1^\dagger (\kappa_{r-1}) & K_1(\kappa_{r-2}, \kappa_{r-1}) & \ldots & 0 & 0 & 0

\end{pmatrix},
\end{eqnarray}

and for even $r=2p$ we have

\begin{eqnarray}
\label{ANeven}
A^{(2p)}=
\begin{pmatrix}
0 & 0 & \ldots & 0 & 0 & Q_1(\kappa_1) \\
0 & 0 & \ldots & 0 & K_2(\kappa_2, \kappa_3) & K_1(\kappa_1, \kappa_2) \\
0 & 0 & \ldots & K_2(\kappa_4, \kappa_5) & K_1(\kappa_3, \kappa_4) & 0 \\
\vdots & \vdots &  & \vdots & \vdots & \vdots  \\
K_2(\kappa_{r-2}, \kappa_{r-1}) &  K_1(\kappa_{r-3}, \kappa_{r-2}) & \ldots & 0 & 0 & 0 \\
Q_2(\kappa_{r-1}) & 0 & \ldots & 0 & 0 & 0
\end{pmatrix}.
\end{eqnarray}
%
Thus, our Hermitian unitary matrix ${\cal S}$, for even $n$, is parametrized by $\frac{n}{2}-1$ complex numbers.
Note that the family of matrices (\ref{ANodd}), (\ref{ANeven}) for $r>3$ 
may not exhaust all possible matrices that
contain $(r-2)^2$ zero elements. For the case $r=4$ the matrix

\begin{eqnarray}
\label{A4}
A^{(4)}_0=\frac{1}{\sqrt{1+|\kappa_1|^2+|\kappa_2|^2}}
\begin{pmatrix}
0 & \kappa_2 & \kappa_1 & 1 \\
\kappa_2 & 0 & 1 & -\kappa_1^* \\
-\kappa_1 & 1 & 0 & -\kappa_2^* \\
1 & \kappa_1^* & -\kappa_2^* & 0
\end{pmatrix}
\end{eqnarray}
is also unitary and contains $(r-2)^2=4$ zero elements, but cannot be obtained from $A^{(4)}$ (\ref{ANeven}) by changing
a numeration of edges and taking particular values of $\kappa_1,\kappa_2$ and $\kappa_3$. However, in Appendix we consider other types of Hermitian unitary
matrices and find out that the matrix structures (\ref{ANodd}), (\ref{ANeven}) and (\ref{A4}) have the 
optimal form to contain maximal number of zeros.

\section{Scattering matrices of quantum star graphs with odd vertex degree} \label{sec4}

In this section we will consider a class of odd dimensional scattering matrices with maximal number of zero. The construction (\ref{S_even}) cannot be
used any more for deriving scattering matrices for $n=2r+1$. Nevertheless, we use a scheme (\ref{ANodd}) by inserting some constrictions
on parameters $\kappa_{i}$ $(i=1,2,...,r-1)$. The matrix $A^{(2p+1)}$ is unitary but not Hermitian for arbitrary values of $\kappa_i$.
If we assume $\kappa_{i}=\kappa_{r-i}$ $(i=1,2,...,r/2)$ and take into account conditions

\begin{equation}
\label{K1K2cond}
\begin{array}{c}
K_1^\dagger(\kappa_1,\kappa_2)=K_1(\kappa_2,\kappa_1),
\\
K_2^\dagger(\kappa_1,\kappa_2)=K_2(\kappa_2,\kappa_1),
\end{array}
\end{equation}
then the matrix (\ref{ANodd}) will become Hermitian. Thus, we retrieve a family of scattering matrices in the form

\begin{equation}
\label{SNodd1}
{\cal S}^{(4p+3)}=
\begin{pmatrix}
0 & 0 & ... & 0 & 0& ... &  0 & Q_1(\kappa_1) \\
0 & 0 & ... & 0 & 0& ... &  K_2(\kappa_2, \kappa_3) & K_1(\kappa_1, \kappa_2) \\
0 & 0 & ... & 0 & 0& ... &  K_1(\kappa_3, \kappa_4) & 0 \\
\vdots & \vdots &  & \vdots & \vdots & & \vdots & \vdots  \\
0 & 0 & ... & 0 & K_2(\kappa_{r-1},\kappa_r) & ... & 0 & 0  \\
0 & 0 & ... & K_2^{\dagger}(\kappa_{r-1}, \kappa_r) & K_1(\kappa_r, \kappa_r) & ... & 0 & 0 \\
\vdots & \vdots &  & \vdots & \vdots & & \vdots & \vdots  \\
0 & K_2^{\dagger}(\kappa_2, \kappa_3) & ... & 0 & 0 & ... & 0 & 0 \\
Q_1^\dagger (\kappa_{1}) & K_1^{\dagger}(\kappa_1, \kappa_2) & ... & 0 & 0 & ... & 0 & 0

\end{pmatrix},
\end{equation}

when $r=2p+1$ and

\begin{equation}
\label{SNodd2}
{\cal S}^{(4p+1)}=
\begin{pmatrix}
0 & 0 & ... & 0 & 0& ... &  0 & Q_1(\kappa_1) \\
0 & 0 & ... & 0 & 0& ... &  K_2(\kappa_2, \kappa_3) & K_1(\kappa_1, \kappa_2) \\
0 & 0 & ... & 0 & 0& ... &  K_1(\kappa_3, \kappa_4) & 0 \\
\vdots & \vdots &  & \vdots & \vdots & & \vdots & \vdots  \\
0 & 0 & ... & K_2(\kappa_r, \kappa_r) & K_1(\kappa_{r-1},\kappa_r) & ... & 0 & 0  \\
0 & 0 & ... & K_1^{\dagger}(\kappa_{r-1}, \kappa_r) & 0 & ... & 0 & 0 \\
\vdots & \vdots &  & \vdots & \vdots & & \vdots & \vdots  \\
0 & K_2^{\dagger}(\kappa_2, \kappa_3) & ... & 0 & 0 & ... & 0 & 0 \\
Q_1^\dagger (\kappa_{1}) & K_1^{\dagger}(\kappa_1, \kappa_2) & ... & 0 & 0 & ... & 0 & 0

\end{pmatrix}
\end{equation}
for even $r=2p$. In Appendix it is shown that the  matrices (\ref{SNodd1}) and (\ref{SNodd2}) contain maximal number of zeros in the class of odd dimensional Hermitian unitary matrices.
They are parametrized by $\frac{n-1}{2}$ complex numbers.

It can be seen that the matrix ${\cal S}^{(4p+3)}$ has $r$ positive eigenvalues $+1$ and
$r+1$ negative eigenvalues $-1$. Unlike ${\cal S}^{(4p+3)}$, the matrix ${\cal S}^{(4p+1)}$ has $r+1$
positive and $r$ negative eigenvalues. To reconstruct boundary conditions (\ref{e1}) for odd $r$, we use the first part of the formula (\ref{e25}).
Taking into account $m=r$ and ${\cal S}_{11}=0$, it follows $T={\cal S}_{12}$. For even $r$ we have $m=r+1$, ${\cal S}_{22}=0$ and using
the second part of (\ref{e25}) it can be noticed that $T={\cal S}_{21}^\dagger={\cal S}_{12}$.

Now let us add a constant potential $U$ on the $(2r+1)$-th edge and consider a scattering problem of particle,
coming from the first line. When we have no potential $(U=0)$, the particle can pass only to $(2r)$-th or $(2r+1)$-th edges.
The transmission probability amplitude into the $(2r)$-th line is 
$\kappa_1$ times higher than the one into the $(2r+1)$-th one. After adding a potential,
these amplitudes are changed and also passing probabilities into other lines appear. To calculate new probability amplitudes,
we use formulae (\ref{SFTpot}, \ref{SFTQpot}) with $V_i=0$ for $i=1,...,2r$ and $V_{2r+1}=U$.
In the case $r=1$ for the first column of scattering matrix we obtain
\begin{eqnarray}
\label{S3}
S_{11}&=&\frac{1-\sqrt{1 - U/E} }{1+\sqrt{1 - U/E}+ 2 \left| \kappa_1 \right|^2}, \nonumber\\
S_{21}&=&\frac{2 \kappa_1^* \sqrt{1+\left| \kappa_1 \right|^2 }}{1+\sqrt{1 - U/E}+ 2 \left| \kappa_1 \right|^2}, \\
S_{31}&=&\frac{2 \left( 1-U/E \right)^{1/4} \sqrt{1+\left| \kappa_1 \right|^2 }}{1+\sqrt{1 - U/E}+ 2 \left| \kappa_1 \right|^2}. \nonumber
\end{eqnarray}

\begin{figure}[ht]\center
\begin{tabular}{ccc}
\includegraphics[width=43mm]{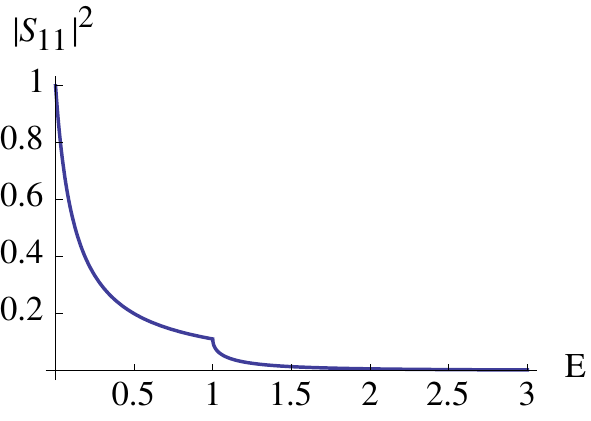}
&
\includegraphics[width=43mm]{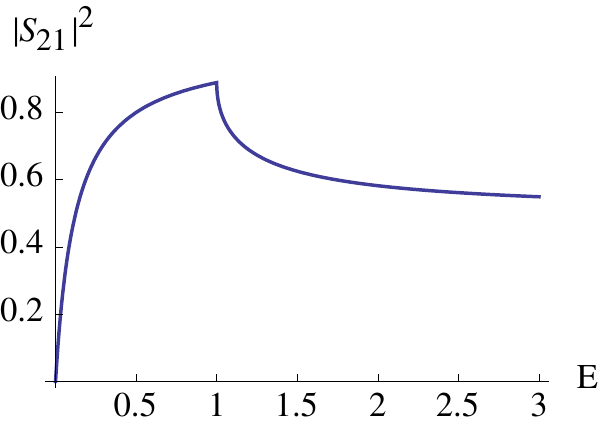}
&
\includegraphics[width=43mm]{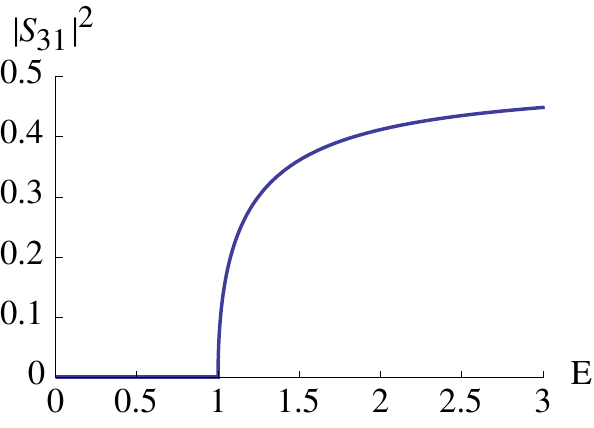} \\
(a)
&
(b)
&
(c)
\end{tabular}
\caption{Energy dependence of reflection probability $\left| S_{11} \right|^2$ (a) and passing probabilities $\left| S_{21} \right|^2$ (b)
and $\left| S_{31} \right|^2$ (c) in the case $U=1$ and $\kappa_1=1$ for the quantum star graph with $3$ edges. }
\label{fig2}
\end{figure}

In figure \ref{fig2} it is shown passing probabilities $\left| S_{i1} \right|^2$ in the case $U=1$ and $\kappa_1=1$,
where it was taken account $S_{31}=0$, when $E<U$.

\begin{figure}[ht]\center
\begin{tabular}{cc}
\includegraphics[width=51mm]{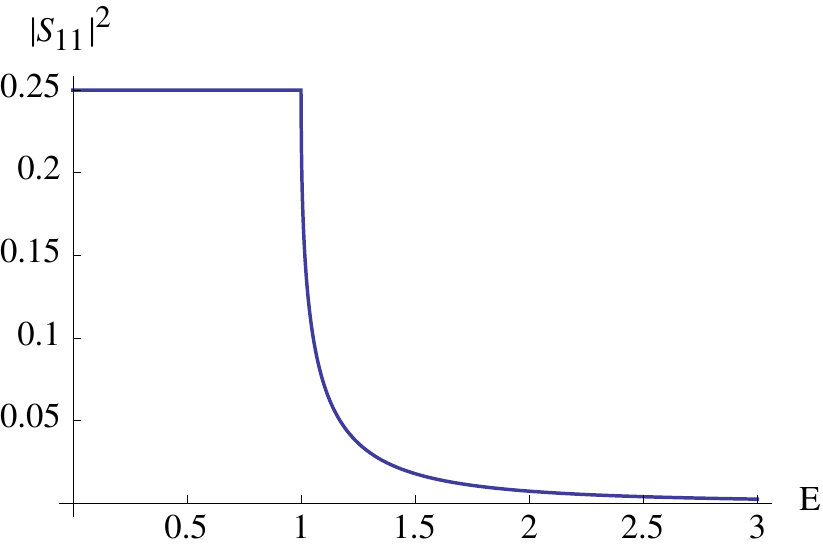}
&
\includegraphics[width=51mm]{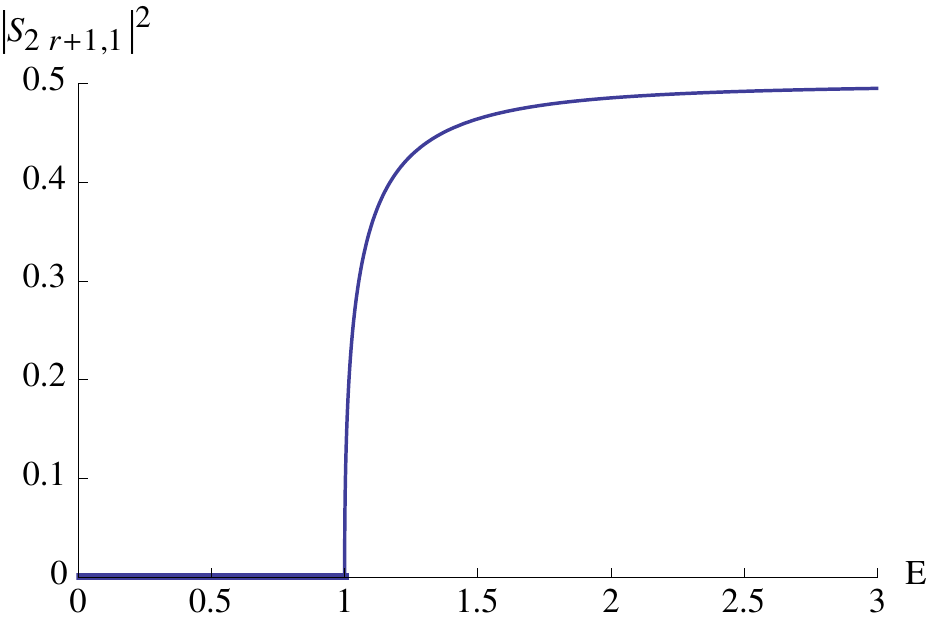} \\
(a)
&
(b)
\end{tabular}
\caption{Energy dependence of reflection probability $|S_{11}|^2$ (a) and passing probability $|S_{2r+1,1}|^2$ (b)
 in the case $U=1$ and $\kappa_1=1$ for the quantum star graph with $2r+1$ edges $(r>1)$.}
\label{fig3}
\end{figure}

For the case $r>1$ calculations show that
\begin{eqnarray}
\label{S579}
S_{11}&=&\frac{-1+\frac{2E}{U}\left(1-\sqrt{1-U/E} \right)}{1+  \left| \kappa_1 \right|^2}, \nonumber\\
S_{21}&=&\frac{\left(1+\frac{2E}{U}\left(1-\sqrt{1-U/E} \right)\right)\kappa_1^*}{\sqrt{1+ \left| \kappa_2 \right|^2}\left(1+  \left| \kappa_1 \right|^2\right)} \nonumber, \\
S_{31}&=&-\frac{\left(-1+\frac{2E}{U}\left(1-\sqrt{1-U/E} \right)\right)\kappa_1^*\kappa_2}{\sqrt{1+ \left| \kappa_2 \right|^2}\left(1+  \left| \kappa_1 \right|^2\right)}, \\
S_{2r,1}&=&\frac{\kappa_1^*}{\sqrt{1+\left|\kappa_1\right|^2}}, \nonumber \\
S_{2r+1,1}&=&\frac{2E\left(1-U/E\right)^{1/4}\left(1-\sqrt{1-U/E}\right)}{U\sqrt{1+\left| \kappa_1\right|^2}}. \nonumber
\end{eqnarray}
Other elements of the first column of new scattering matrix are zeros. In figure \ref{fig3} it is shown an energy dependence of reflection
probability $\left|S_{11}\right|^2$ (fig. \ref{fig3}a)  and passing probability $\left|S_{2r+1,1}\right|^2$ (fig. \ref{fig3}b).
The passing probabilities $\left|S_{21}\right|^2$ and $\left|S_{31}\right|^2$ can be obtained from figure \ref{fig3}a by
multiplying with appropriate constants. Thus, we find out that in the case $E<U$ the reflection probability and
passing probabilities of particle to other edges don't depend on energy.

\section{Physical realization of quantum vertices with minimal passbands}
 
To apply all useful properties of singular quantum vertices in practice, it is important to design realization schemes of such "exotic" boundary conditions.
It is well-known that the simplest case is a $\delta$ potential, which can be achieved if one maintains the product of the width of the well and the strength
of potential constant while decreasing the well's width and increasing a potential. The next important case is a $\delta'$ potential which can be reached
by using three $\delta$ potentials with renormalized strengths in the disappearing distances \cite{delta-prime, ENZ01}. More general point interactions with 
scale invariant type couplings
can be approximated by supplementary graphs with $\delta$ vertices. Also in some cases the localized magnetic fields have to be applied on edges to attain phase
changes. We are going to apply the following construction scheme: \cite{Inverse-prob}
 
 \begin{enumerate}
\item Let us disconnect $n$ edges of the graph and reconnect endpoints $\{i, j\}\; (i\neq j)$ by additional edges of lengths ${d}/{r_{ij}}$. In the case $r_{ij}=0$
we will assume that the pair remains unconnected. To produce an extra phase shift $\chi_{ij}$ between connected endpoints $\{i, j\}$, we apply a vector potential
$A_{ij}$ on the line $(i,j)$. Let us place a $\delta$ coupling of strength $v_i$ at each endpoint $i$.

\item The length ratios $r_{ij}$ and the phase shifts $\chi_{ij}$ are defined from the matrix
\begin{eqnarray}
\label{efcnd1}
Q = \begin{pmatrix} T \\ I^{(n-m)}\end{pmatrix} \begin{pmatrix} -T^\dagger & I^{(n-m)}\end{pmatrix}
= \begin{pmatrix} -T T^\dagger & T \\ -T^\dagger & I^{(n-m)}\end{pmatrix}
 \end{eqnarray}
using the relation $r_{ij} e^{{\rm i}\chi_{ij}}=Q_{ij}$ ($i \ne j$). 

\item The $\delta$ potential strengths $v_{i}$ is determined by the diagonal elements of the matrix
\begin{eqnarray}
 \label{vicnd}
V = \frac{1}{d} (2 I^{(n)}-J^{(n)}) R,
\end{eqnarray}
where the matrix $R$ is formed from the absolute values of the matrix elements $Q_{ij}$, i.e., $R=\{r_{ij}\} =\{|Q_{ij}|\}$; and the
$n \times n$ matrix $J^{(n)}$ has all elements equal to one.

\end{enumerate}

For boundary conditions obtained in previous sections in both cases of odd and even $n$, the $T$ matrix is unitary, which means
 
\begin{eqnarray}
\label{efcnd}
Q = \begin{pmatrix} -I^{(m)} & T \\ -T^\dagger & I^{(n-m)}\end{pmatrix}.
 \end{eqnarray}
Thus, it can be seen that our construction represents a bipartite graph. The corresponding scheme of boundary condition (\ref{e1N})
consists of two disjoint sets, each containing $n/2$ points. Distances and phase differences between vertices are determined
by $r_{ij}=|A_{i,j-n/2}|$ and $\chi_{ij}=Arg(A_{i,j-n/2})$ for $1\leq i \leq n/2, n/2+1 \leq j \leq n$ and other vertices are disconnected.
For $\delta$ potential strengths we have

\begin{figure}[ht]\center
\begin{tabular}{ccc}
\includegraphics[width=41mm]{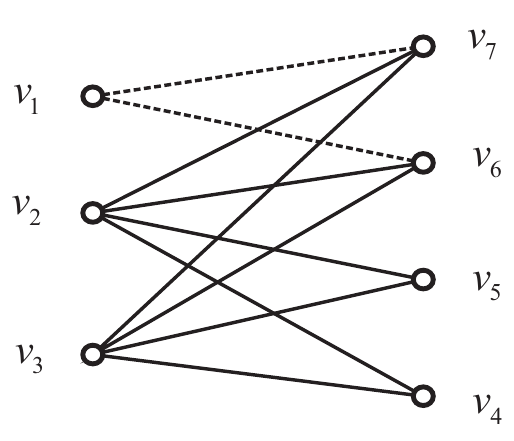}
&
\includegraphics[width=41mm]{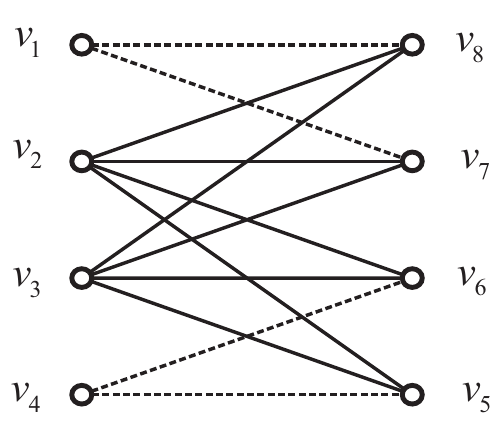}
&
\includegraphics[width=41mm]{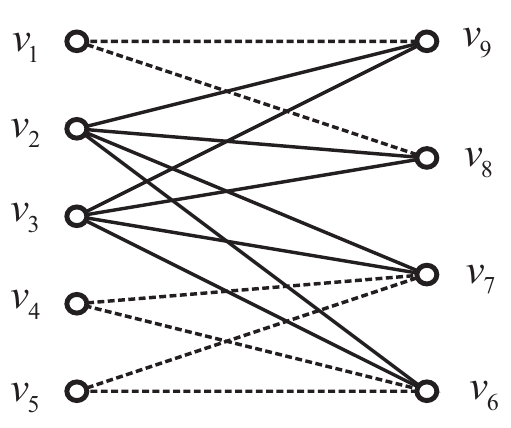} \\
(a)
&
(b)
&
(c)
\end{tabular}
\caption{Finite approximations of quantum vertex with minimal number of passbands for different value of vertex degree: $n=7$ (a), $n=8$ (b) and $n=9$ (c). }
\label{fig4}
\end{figure}

\begin{eqnarray}
v_i&=&\frac{1}{d}\left(1- \sum_{ j=1}^{n/2}|A_{ij}|\right),\; 1\leq i\leq n/2, \nonumber\\
v_i&=&\frac{1}{d}\left(1- \sum_{ j=1}^{n/2}|A_{j,i-n/2}|\right), \; n/2 +1\leq i\leq n.
\label{vi}
\end{eqnarray}
As an example let us describe construction schemes corresponding to scattering matrices with maximal number of zeros in the case
$\kappa_i=1$ for $i=1,\ldots, n$. In this particular case the all matrix elements of scattering matrix are real, therefore all phase differences vanish.
In figure \ref{fig4} there are shown disjoint sets of our construction for various cases of vertex degree of initial graph. For odd dimensions in
the case $n=4p+3$ the first disjoint set contains $\frac{n-1}{2}$ vertices, in the case $n=4p+1$ it has $\frac{n+1}{2}$ ones. 
For both even and odd dimensions the bipartite graph contains two type of edges. In figure \ref{fig4}a we have $r_{16}=r_{17}=\frac{1}{\sqrt{2}}$
and other length ratios of connected lines are equal to $\frac{1}{2}$. For potential strengths we have $v_1=\frac{1-\sqrt{2}}{d}$, $v_2=v_3=-\frac{1}{d}$, $v_4=v_5=0$,
and $v_6=v_7=-\frac{1}{\sqrt{2}d}$. In figure \ref{fig4}b $r_{17}=r_{18}=r_{45}=r_{46}=\frac{1}{\sqrt{2}}$ and other length ratios are
$\frac{1}{2}$. Potential strengths are $v_1=v_4=\frac{1-\sqrt{2}}{d}$, $v_2=v_3=-\frac{1}{d}$ and $v_5=v_6=v_7=v_8=-\frac{1}{\sqrt{2}d}$.
It is also not hard to calculate length ratios and potential strengths for figure \ref{fig4}c: $r_{18}=r_{19}=r_{46}=r_{47}=r_{56}=r_{57}=\frac{1}{\sqrt{2}}$,
$r_{26}=r_{27}=r_{28}=r_{29}=r_{36}=r_{37}=r_{38}=r_{39}=\frac{1}{2}$, $v_1=\frac{1-\sqrt{2}}{d}$, $v_2=v_3=v_6=v_7=-\frac{1}{d}$, $v_5=v_6=0$, 
$v_9=-\frac{1}{\sqrt{2}d}$.

\section{Conlcusion}

We considered a class of boundary conditions of scattering problem of quantum graph with two important properties.
On the one hand the scattering matrix contains maximal number of zeros, on the other hand such type of vertices show universal
filtering properties. The limited number of nonzero elements in scattering matrices restricts passing directions of incoming
particle. After reconstruction of boundary condition from scattering matrix, we added a constant controlling potential
to the last edge, which affects on elements of scattering matrices. We found that for quantum graph with vertex order $n>3$
passing and reflecting probabilities of particle with lower energy than potential strength, initially coming from the first edge, are independent
from particle's energy.

\medskip
We acknowledge the  support  by the Japan Ministry of Education, Culture, Sports, Science and Technology under the Grant number 24540412.


\appendix

%
\section{Proof of proposition about maximal number of zeros}

In this appendix we will show that the Hermitian unitary matrix of dimension $n$ can contain no more than $(n-2)^2$ zeros for odd $n$
and $(n-2)^2+4$ zeros for even $n$. Note that we consider only scattering matrices corresponding to completely connected vertices.
In this proof we will discuss different cases of matrix structures and show that the scattering matrices introduced in sections \ref{sec3} and \ref{sec4}
contain maximal number of zeros. Since the whole proof is fairly long, we will skip some details. However, more elegant and
explicit proof will be appreciated.

Let us simultaneously permute rows and columns of scattering matrix which just corresponds to renumeration of graph edges. If some block of size $p\times p\; (p<n/2)$ appears
on the upper right corner of the matrix then from the condition of Hermiticity it follows that we have a complex conjugate block on the lower left corner.
It can be seen that the first $p$ edges of the graph are connected with last $p$ lines and the rest $n-2p$ edges are separated from them.
Hence, for odd dimensions the scattering matrix cannot be reduced to blocked form and for even dimensions we can have only blocks of
size $n/2$ as in the form (\ref{S_even}). Also it can be noticed that the blocks of size $p\times q$, $p \neq q$ are not allowed since assuming
$p>q$ $(p<q)$ it follows that we have an orthogonal set of $q$-dimensional ($p$-dimensional)  $p$ rows ($q$ columns). 

Now let us assume that we found an Hermitian unitary matrix of size $n \times n$ and consider the row containing maximal number of zeros.
It has to contain at least one non-zero element as the sum of squared modules of row elements is equal to $1$. However, if the row contains
only one non-zero element, it corresponds to splitted graph into two independent parts. If that element is non-diagonal then
we have two connected lines without any interaction. In the case of diagonal non-zero element we have a single line with reflection probability $1$.

Suppose we have a row with two non-zero elements and $n-2$ zeros. By swapping simultaneously rows and columns it is always possible to turn
them to upper right corner of the matrix, i.e., to make $S_{1,\kappa}=0$ for $\kappa=1,\ldots, n-2$ and $S_{1,n-1}\neq 0$, $S_{1,n}\neq 0$. Now let us consider
$(n-1)$-th and $n$-th columns and suppose $S_{j_1,n} \neq 0$ for some $1<j_1\leq n$. From the orthogonality condition of the first and $j_1$-th rows it follows
$S_{j_1,n-1} \neq 0$. After interchanging the second and $j_1$-th rows and columns of the ${\cal S}$-matrix, we will obtain $S_{2,n-1}\neq 0, S_{2,n}\neq 0$. Assume that
$S_{2,\kappa}=0$ for all $\kappa=1,\ldots, n-2$. Then it can be noticed that $S_{n-1,\kappa}=0, S_{n,\kappa}=0$ for $\kappa=2,\ldots, n$ since it is impossible to find more than two 
non-zero vectors, which are orthogonal to each other. Thus, we obtained a block of size $2\times 2$ which means that the first, second,
$(n-1)$-th and $n$-th lines are effectively disconnected from other part of graph. Therefore the second row has to contain at least one more non-zero
element.

Assume that  $S_{2,\kappa}=0$ for $\kappa=1,\ldots, n-3$ and $S_{2,j}\neq 0$ for $j=n-2, n-1, n$. Then from the orthogonality of columns and rows it can be concluded that
there is some $j_2>2$ for which $S_{j_2,n-2}\neq 0, S_{j_2,n-1}\neq 0$ and $S_{j_2,n}\neq 0$. By interchanging rows and columns it is always possible to make $j_2=3$.
If $S_{3,\kappa}=0$ for $\kappa=1,\ldots, n-3$ then $S_{i,\kappa}=0$ for $i=4,\dots n$, $\kappa=1,2,3$, i.e., we have a separated block.
From unitarity condition of $S$-matrix it follows also that matrix elements of such block can be parametrized by
no more than two constants. Hence, using parameters $\kappa_1$ and $\kappa_2$ we obtain a matrix block (\ref{A3}), which can be used to form
$S$-matrix with maximal number of zeros in the case $n=6$. Adding a condition of Hermiticity we will have a reduction of one parameter and by substituting $\kappa_1=\kappa_2$
we come to the matrix $S^{(3)}$ (\ref{SNodd1}). Let us consider the next case, which is $S_{3,\kappa}\neq 0, \kappa=n-3, \ldots ,n$.
Applying orthogonality condition of rows and columns it can be proved that there is a row $j_3>3$ for which
$S_{j_3,\kappa}\neq 0$, $\kappa=n-3,\ldots, n$. Swapping rows and columns we can make $j_3=4$ and it is not hard to see that further construction doesn't give an
optimal matrix form containing maximal number of zeros comparing to matrices of sections \ref{sec3} and \ref{sec4} .

Let us consider the next case, which is $S_{2,j}\neq 0$, $j=n-3,\ldots , n$ and $S_{2,\kappa}=0$, $\kappa=1,\ldots, n-4$. Here also
we can find some $j_4>2$ for which $S_{j_4,n-2}\neq 0, S_{j_4,n-1}\neq 0, S_{j_4,n}\neq 0$ and set $j_4=3$. It is obvious that if $S_{3,\kappa}\neq 0, \kappa=1,\ldots, n-3$
then the scattering matrix can be reduced to the previous case. Thus, assume that some $j_5<n-2$ exists for which $S_{3,j_5}\neq 0$. At first consider the case
$j_5<n-3$ and $S_{3,n-3}=0$. Setting $j_5=n-4$ it can be proved that $S_{4,\kappa}\neq 0$ for $\kappa=n-4,\ldots,n$, which eventually brings to a matrix form
containing less zeros than the matrices of sections \ref{sec3} and \ref{sec4}. Let us consider the case $j_5=n-3$, which means $S_{3,\kappa}\neq 0, \kappa=1,\ldots, n-4$ and $S_{3,n-3}\neq 0$.
From orthogonality condition of rows and columns it can be proved the existence of some $j_6>3$ for which $S_{j_6,n-3}\neq 0$, $S_{j_6,n-2}\neq 0$
and assume $j_6=4$. As we are searching a matrix form with maximal number of zeros, we will take $S_{4,n-1}=0$, $S_{4,n}=0$. We notice that we
derived a structure of matrix (\ref{ANeven}) for the case $r=4$. Using similar strategy for searching optimal matrix form containing maximal
number of zeros, we will come to matrices (\ref{ANodd}) and (\ref{ANeven}) for odd and even $r$ respectively. 

Note that from unitarity condition it is possible to prove that our matrix constructions can be parametrized by $r-1$ complex constants.
In the case of even vertex degree $n$ these matrices can be used to form a block $A$ of the scattering matrix (\ref{S_even}).
In the case of odd $n$ the scattering matrices cannot contain blocks, therefore the
condition of Hermiticty restricts a number of parameters and we will have $\frac{n-1}{2}$ complex parameters.

Now let us consider the case when the scattering matrix doesn't contain any row with two non-zero elements. It is implied that 
there are at least four rows, containing only three non-zero elements. We will show that if we try to add one more row with three non-zero
elements, the corresponding row with five non-zero elements will appear. Assume that $S_{1,\kappa}=0$ for $\kappa=1,\ldots, n-3$ and
$S_{1,n-2}\neq 0$, $S_{1,n-1}\neq 0$, $S_{1,n}\neq 0$.

Consider a case when there is no such row $j_7>1$ for which in the same time
we will have $S_{j_7,n-2}\neq 0$, $S_{j_7,n-1}\neq 0$, $S_{j_7,n}\neq 0$. Then from orthogonality of rows we can always find rows
$j_8$, $j_9$ and $j_{10}$ for which $S_{j_8,n}\neq 0$, $S_{j_8,n-1}\neq 0$, $S_{j_9,n}\neq 0$, $S_{j_9,n-2}\neq 0$, $S_{j_{10},n-1}\neq 0$,
$S_{j_{10},n-2}\neq 0$ and assume $j_8=2, j_9=3, j_{10}=4$. It can be seen that there is a column $j_{11}<n-2$ with $S_{2,j_{11}}\neq 0$, $S_{3,j_{11}}\neq 0$,
$S_{4,j_{11}}\neq 0$ and let us assume $j_{11}=n-3$. Thus, we derive a matrix form (\ref{A4}), which can be used to construct a Hermitian unitary matrix
of dimension $n=8$. Using a condition of unitarity of ${\cal S}$-matrix it is not hard to prove that the whole set of matrices of such form can be parametrized by two
complex parameters as in (\ref{A4}). It can be shown also that if we try to build larger matrices based on this construction we have to add some non-zero elements
on second, third and forth rows, which will bring us to non-optimal matrix structure for containing maximal number of zeros.

Let us consider the opposite case, i.e., there is a row $j_7>1$ for which $S_{j_7,n-2}\neq 0$, $S_{j_7,n-1}\neq 0$, $S_{j_7,n}\neq 0$ and put $j_7=2$.
It is reasonable to take $S_{2,\kappa}=0$ for $\kappa=1,\ldots, n-3$, else it will be always possible to find a corresponding row, containing at least five non-zero elements.
Then we can accept $S_{3,n-3}\neq 0$, $S_{3,n-2}\neq 0$, $S_{3,n-1}\neq 0$, $S_{4,n-3}\neq 0$, $S_{4,n-2}\neq 0$, $S_{4,n-1}\neq 0$. If we take
$S_{3,n}=0$, $S_{4,n}=0$ we will come to a transposed form of the matrix (\ref{ANeven}) in the case $r=4$. If we try to construct unitary Hermitian matrices
of larger dimension with maximal number of zeros, we eventually derive a transposed form of the (\ref{ANeven}).


%

\end{document}